\def\be{\begin{equation}}
\def\ee{\end{equation}}
\def\ber{\begin{eqnarray}}
\def\eer{\end{eqnarray}}
\def\bers{\begin{eqnarray*}}
\def\eers{\end{eqnarray*}}

\def\etal{{\it et al.}\/}

\def\mahg{mAhg$^{-1}$}

\documentclass[aps,prb,twocolumn,showpacs,groupedaddressamsmath,amssymb,floatfix,superscriptaddress]{revtex4-1}
\usepackage{dcolumn}   % Align table columns on decimal point
\usepackage{amsmath}
\usepackage{graphicx}    % Include figure files
\usepackage{subfigure}
\usepackage{color}
\newcommand{\comment}[1]{}
\usepackage{ifthen}
\newboolean{includefigs}
\setboolean{includefigs}{true}      % Permits rapid compiling by removing figures/text
\newboolean{includetext}
\setboolean{includetext}{true}     % false = removed, true = included
\newcommand{\condcomment}[2]{\ifthenelse{#1}{#2}{}}

\begin{document}
%%%%%%%%%%%%%%%%%%%%%%%%%%%%%%%%%%%%%%%%%%%%%%%%%
%               TITLE 
%%%%%%%%%%%%%%%%%%%%%%%%%%%%%%%%%%%%%%%%%%%%%%%%%
\title{ Enhanced Li capacity in functionalized graphene: A first principle study with van der Waals correction}
\author{Rajiv K. Chouhan}\affiliation{ Department of Physics, Boise State University, 1910 University Dr., Boise, Idaho 83725, USA}
\email{rajivchouhan@boisestate.edu}
\author{Pushpa Raghani}\affiliation{ Department of Physics, Boise State University, 1910 University Dr., Boise, Idaho 83725, USA}

%%%%%%%%%%%%%%%%%%%%%%%%%%%%%%%%%%%%%%%%%%%%%%%%%
%               ABSTRACT 
%%%%%%%%%%%%%%%%%%%%%%%%%%%%%%%%%%%%%%%%%%%%%%%%%
\begin{abstract}
We have investigated the adsorption of Li on graphene oxide using density functional theory. We show a novel and simple approach to achieve a positive lithiation potential on epoxy and hydroxyl functionalized graphene, compared to the negative lithiation potential that has been found on prestine graphene. We included the van der Waals correction into the calculation so as to get better picture of weak interactions. A positive lithiation potential suggests a favorable adsorption of Li on graphene oxide sheets that can lead to an increase in the specific capacity, which in turn can be used as an anode material in Li-batteries. We find a high specific capacity of $\sim$860 \mahg by functionalizing the graphene sheet. This capacity is higher than the previously reported capacities that were achieved on graphene with high concentration of Stone-Wales (75\%) and divacancy (16\%) defects. Creating such high density of defects can make the entire system energetically unstable, whereas graphene oxide is a naturally occurring substance.
\end{abstract} 
\date{\today}
\maketitle

\section{Introduction}

Energy storage is a major challenge in present times to meet the demand of consistent power supply, both in the portable devices like laptops, cameras, cellphone, etc., and in the bigger devices like electric vehicles, backup inverters, etc. Additionally, increase in the cost of fuel and increase in green house gases have given rise to the hybrid electric vehicle technology with battery supplement, known as plugin hybrid electric vehicle (PHEV).\cite{howel2008} For several years, PHEVs and many other appliances, Lithium batteries have been considered to be optimum candidates for rechargeable battery technology.

Although Li metal has a high theoretical capacity of $\sim$3842 \mahg, rechargeable lithium batteries cannot use the bulk lithium metal due to safety issues, dendritic growth, rapid decay and small recycling time.\cite{ref2,muzushima1980,tarascon2001,bhattacharya2010,orsini1998} To overcome these problems many other materials and structure have been investigated, e.g., metal oxide such as LiCoO$_2$, LiFePO$_4$ and LiMnO$_2$,\cite{tarascon2001,muzushima1980,bhattacharya2010,orsini1998} where Li atoms can be stored in layers within these metal oxides. However, these materials are found to have very small practical capacities of 140 \mahg, 170 \mahg and 119 \mahg, respectively. Whereas, graphite with theoretical capacity of $\sim$372 \mahg is used as conventional anode. Recently, significant work is being done to enhance the capacities of anodes as well as cathodes using the carbon based materials,\cite{mukherjee2014,Liu2012,Zhou2012,Krishnan2011,Stournara2011,Meunier2002,dutta2014} like a composite anode based on carbon and silicon with a specific capacity of $\sim$1000 \mahg has been proposed.\cite{Liu2012} Similarly, a few more works\cite{mukherjee2014,dutta2014} also shows a hike in the capacity of the Li storage on graphene porous networks with Stone-Wales (SW) and divacancy (DV) defects. They reported an experimental capacity of 850 \mahg , whereas theoretical calculation predict capacities of 698 \mahg and 590 \mahg with 75\% SW and 16\% DV defects, respectively. Although even higher capacities with 100\% SW and 25\% DV defects are predicted from the calculation, creating graphene sheets with such a high percentage of defects is difficult and energetically unfavourable. For example, formation energies of SW defect is $\sim$5 eV\cite{li2005,ma2009,banhart2011} and it is formed by rapid quenching from high temperature under the radiation of 90-100 eV of electron energy. Similarly, formation energies of single- and di-vacancies in graphene sheet are $\sim$7.5 eV and $\sim$8 eV, respectively.\cite{krasheninnikov2006,rossato2005,barbary2003} Although DV  are thermodynamically more stable than single vacancy at high temperature, these vacancies are practically immobile.\cite{barbary2003} Additionally, removal of a large number of atoms can lead to the bending and wrapping of the sheet destroying its electronic properties. To overcome the problem, as well as taking the advantage of the fact that graphene sheet with non-stoichiometric nature can increase the lithiation potential (LP), we investigated the lithium adsorption over graphene oxide (GO). Interestingly, we find a positive LP on these functionalized graphene sheets suggesting a favourable adsorption of Li on GO leading to the specific capacity of $\sim$860 \mahg. The most notable aspect of our proposed system is that it is simple and cost effective to prepare. Functionalized graphene can be created by simply treating it with strong acid/base\cite{loh2010,suk2010,hummers1958,marcano2010,dreyer2010} that results in functional groups (epoxy (-O-) and hydroxyl (-OH)) attached to the sheet. For GO, one does not have to go through the reduction process of removing oxygen that naturally gets adsorbed on it or one does not have to create any high energy defects. Functionalized graphene shows a great potential for Li storage for making high capacity batteries. A specific capacity of $\sim$860 \mahg obtained by us is greater than that of 698 \mahg and 590 \mahg obtained with 75\% SW and 16\% DV of defects. Thus, GO is not only easy to prepare, it also gives very high specific capacities and could emerge as an ideal material for Li-batteries.

\section{Methods}
Density functional theory (DFT) calculations are performed in the pseudopotential formalism using Quantum Espresso\cite{QE} code. Due to the presence of lithium metal, a higher plane-wave cutoffs of 60 Ry is used. Exchange interactions were treated within GGA with Perdew-Burke-Ernzerhof (PBE)\cite{perdew1996} functional form. We did include spin polarization in our calculations but found that there is negligible magnetic moment in the system. As van-der-waal's interactions play an important role in binding of Lithium ions over the graphene sheet,\cite{Allouche2012} we have included dispersion correction in our DFT calculations.\cite{grimme2006,tkatchenko2009} We choose a computationally cost-effective ``DFT-D2" method of Grimme,\cite{grimme2006} for VdW interactions. For (1$\times$1) unit cell of graphene, we use a Monkhorst-pack k-mesh grid of $18\times18\times1$ and the corresponding k-mesh for bigger supercells. All the configurations have been relaxed with a force tolerance of $10^{-3}$ Ry/bohr. To avoid interactions between images perpendicular to the plane of the sheet, a vacuum of 21 \AA \ is used. Using these parameters, we get C-C bond length of 1.42 \AA \ for pristine graphene that agrees well with the experiments.\cite{trucano1975}

We calculate the adsorption energy and the lithiation potential as a function of Li concentration in pristine and functionalized graphene. LP can be determined from the Nernst equation using following equation:
 \begin{equation}
 LP = \frac{\Delta G_f}{zF},
 \end{equation} where $z$ is charge on the Li ion and $ \Delta G_f= \Delta E_f+T \Delta S_f - P \Delta V_f $ is the change in Gibb's free energy. At room temperature, the entropic effects ($T \Delta S_f$) and volume effects ($P \Delta V_f$) will be very small compared\cite{Aydinol1996} to the formation energy $\Delta E_f$ and can be neglected.   $\Delta E_f$ is defined as $\Delta E_f = -[E_{Li_n+subs} - (nE_{Li}+ E_{subs})]$, where $E_{Li_n+subs}$ is the total energy of the lithiated GO, $E_{Li}$ is the total energy of a single lithium atom in body centred cubic (BCC) structure, and $E_{subs}$ is the total energy of substrate GO. On the other side, the AE of Li is calculated by using $E_{Li}$ for a gas phase Li atom instead of the metallic Li in the BCC structure.

\section{Results and discussion}
\subsection{Adsorption of Li on graphene}
\begin{figure}[t]
\centering
\includegraphics[scale=0.33]{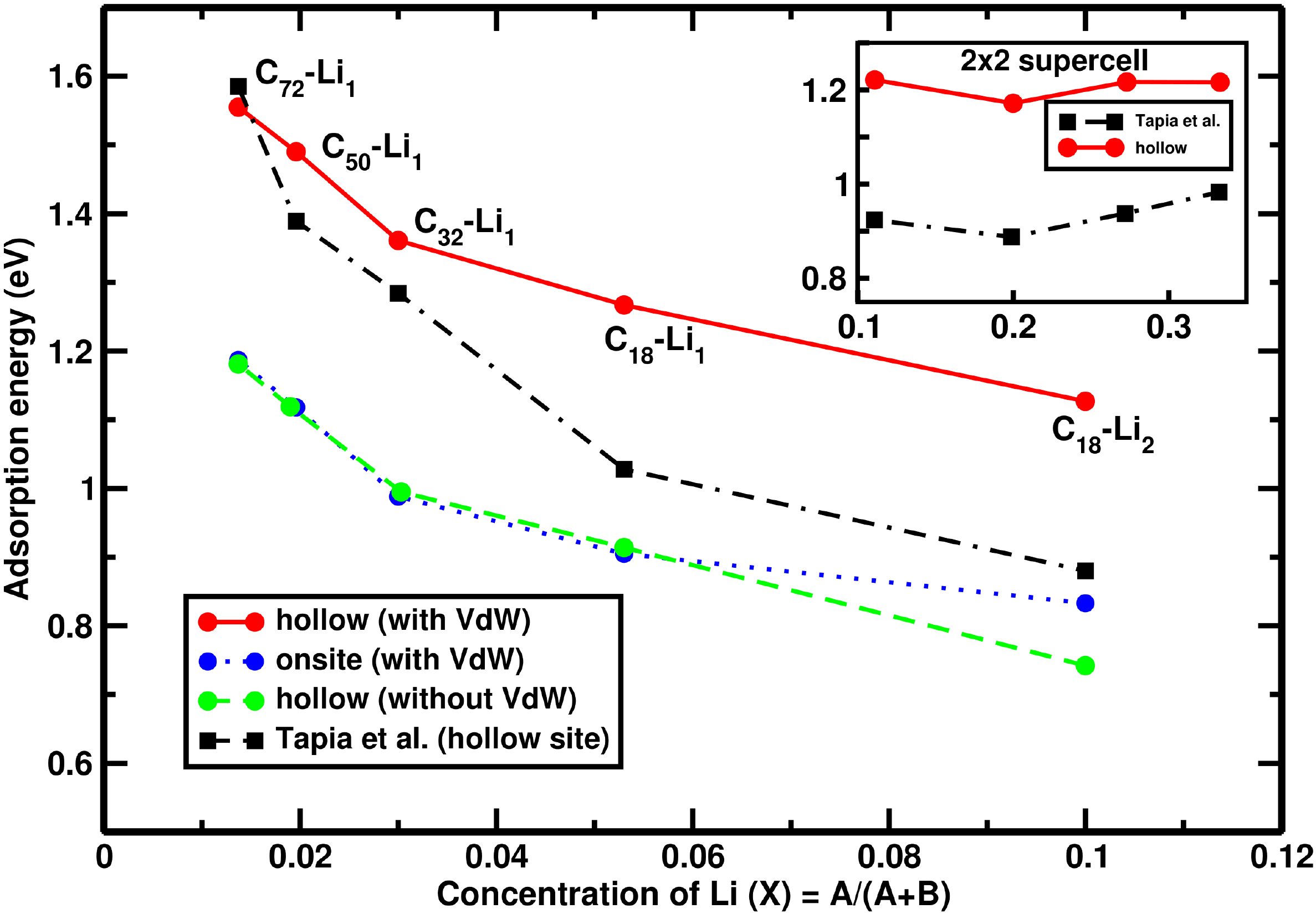}
\caption { Adsorption energies versus Li concentration from 0.01\% to 0.1\% on graphene by varying cell size for hollow and onsite positions. The inset shows adsorption energies versus Li concentrations from 0.1\% to 0.33\% in a fixed cell size of $2\times2$ at hollow site.}
\label{fig1}
\end{figure}

We have looked at the adsorption of Li ion on graphene for various Li concentrations by either varying the cell size or keeping the cell size fixed and varying the number of Li atoms in the cell. Figure [\ref{fig1}] shows the adsorption energies as a function of Li concentration from 0.01\% to 0.10\% by varying the cell size from  3$\times$3 to  6$\times$6, while the inset shows the variation of Li concentration from 0.11\% to 0.33\% by varying number of Li atoms in a fixed cell size  2$\times$2. First of all, notice that adsorption energy (AE) of Li increases with decrease in Li concentration suggesting a repulsive interaction between Li atoms. AE at hollow site is the highest of all the other high symmetry sites (onsite, hollow and bridge), in agreement with previous literature.\cite{chan2008, Garay-Tapia2012} We find that including van-der-waal's (VdW) increases the adsorption of Li by 0.35 eV irrespective of the Li concentration and the dispersion contribution to the AE comes out to be 0.48 eV, in good agreement with 0.45 eV found by Allouche \etal \cite{Allouche2012} Even though the AE changes by a fixed amount of $\sim$0.35 eV for all the Li concentrations, this difference is expected to change when we functionalize graphene with epoxy/hydroxyl groups, as these functional groups will change the charge distribution in the system. Hence, just by taking the VdW contribution separately will not give you the exact information about the AE and the proper inclusion of ``DFT-D2" inside the scf calculation is important.

\subsection{Lithiation of Graphene Oxide}
Here we investigate GO as an anode material for Li-batteries. It is well known that pristine graphene is not a suitable candidate for the Li adsorption due to its negative lithiation potential,\cite{mukherjee2014,Zhou2012_2} and hence it has a limitation to intercalate only between the spaces of graphene planes (with a theoretical capacity of $\sim$372 \mahg). Recently it has been shown that by introducing defects/vacancies,\cite{mukherjee2014,dutta2014,Zhou2012} and functionalizing\cite{Stournara2011,Robledo2014} the graphene sheet, a positive LP can be achieved, and hence one can enhance the battery capacity.

\begin{figure}[h]
\centering
\includegraphics[width=84mm, height=62mm]{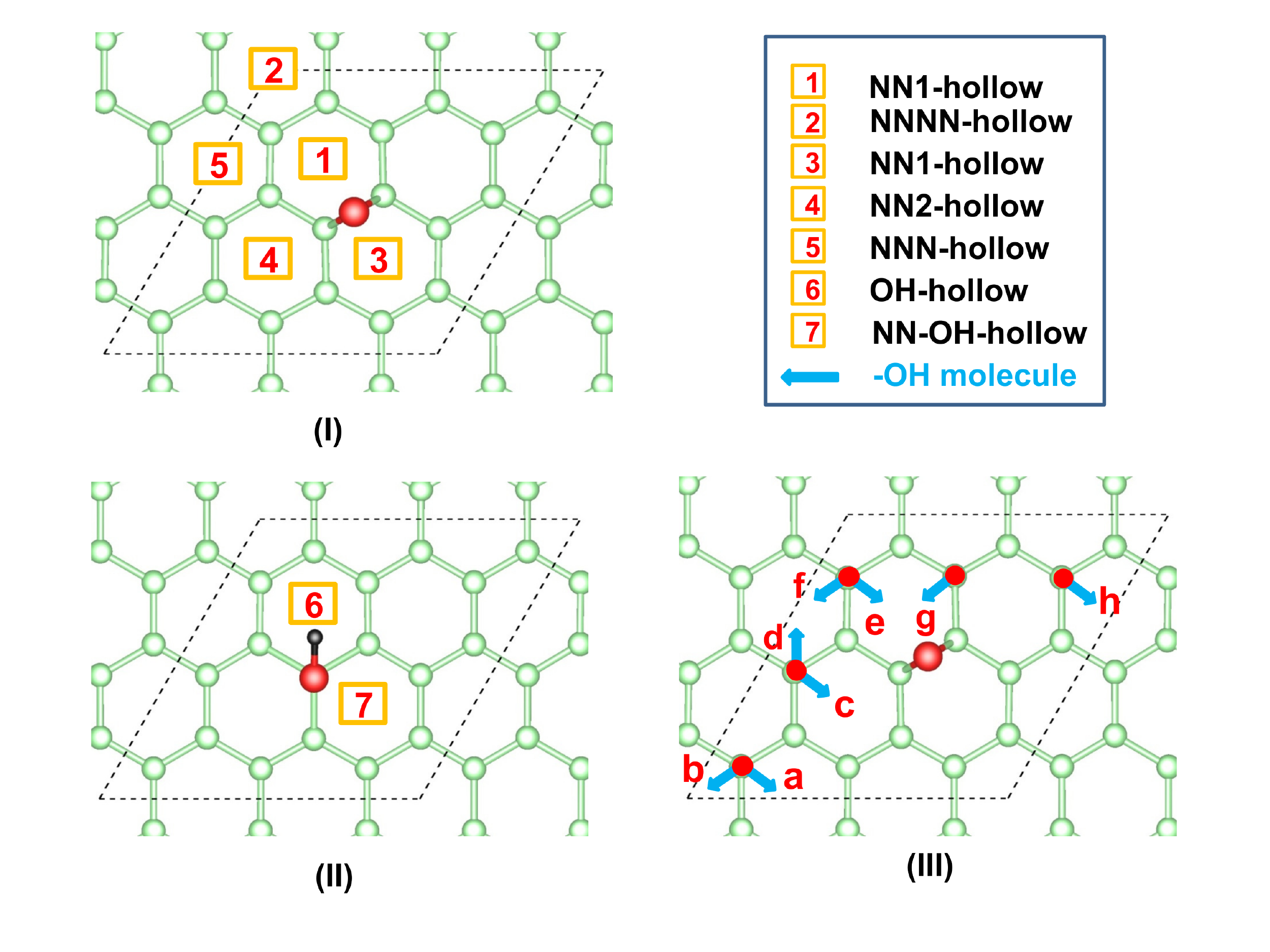}
\caption {Models of GO with three different functional groups: (I) Epoxy (II) Hydroxyl and (III) Epoxy+Hydroxyl, attached to graphene sheet. Squares with numbers show the adsorption sites of Li. In model(III) blue arrows show the possible configurations (a-h) with different orientations on the same side of epoxy functional.}
\label{fig3}
\end{figure}

The structure of GO is still a puzzle due to its non-stoichiometric nature. Most recent NMR experiments\cite{cai2008} suggest that the Carbon:Epoxy:Hydroxyl content on graphene can be varied by changing the  environmental conditions. Commonly, the C:O ratio in GO is found to be  2:1, however after reduction process, this ratio can be increased to as high a value as 14.9:1.\cite{dreyer2010,zhou2013,staudenmaier1898, stanko2007, wang2009} In our work, we use C:O ratio as 18:1 and 9:1. Under O-rich environment, the GO contains epoxy functional group (-O- attached to bridge site), while in H-rich environment, -OH group is formed (attached to carbon atom with orientation along hollow site) as shown in Figure [\ref{fig3}] (I) and (II), respectively. We looked at the Lithiation potential on three types of functionalized graphene: (I) epoxy covered, (II) -OH covered, and (III) both, epoxy and -OH covered as shown in Figure [\ref{fig3}]. These calculation are performed in a 3$\times$3 supercell. Numbers in the Figure[\ref{fig3}] show the positions of Li atoms with respect to the functional group. Arrows in Figure[\ref{fig3}] (III) a-h, indicate possible -OH positions and orientations with respect to the position of the epoxy group in same side as the epoxy group.

\begin{figure}[h]
\centering
\includegraphics[scale=0.046]{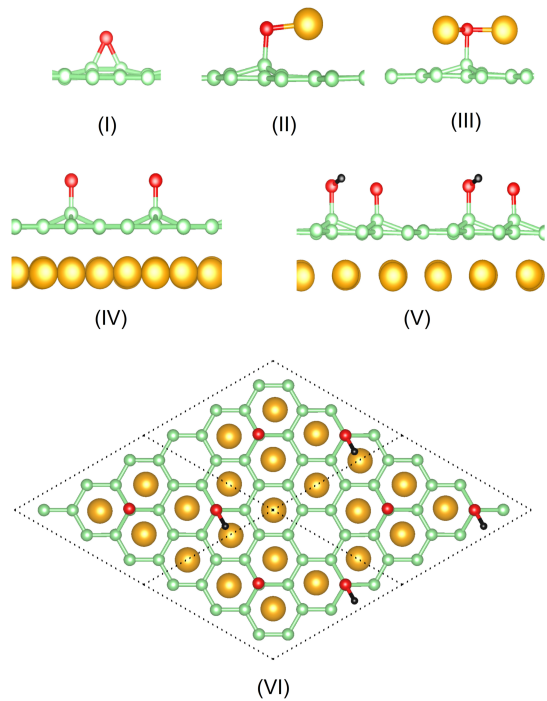}
\caption {Configurations (I)-(IV) show different forms of oxygen obtained on GO sheet in presence of Li. Configuration (V) and (VI) show side and top view of the highest capacity (860 \mahg) configuration with positive lithiation potential.}
\label{fig32}
\end{figure}

For the case of single Li atom adsorption on the opposite side of epoxy group (site-3,4,5), we get the highest LP of 0.03 eV at site-3 having specific capacity of 115 \mahg. For the adsorption at site-1 and 2 on the same side as epoxy group, we get a LP of 1.44 eV and 0.31 eV, respectively. This high LP of 1.44 eV at site-1 is due to the binding of Li atom to the Oxygen atom causing it to break the double bond with the graphene sheet (see Figure[\ref{fig32}](II)). A high LP of 1.66 eV was also found by Stournara \etal,\cite{Stournara2011} for a similar configuration of Li bonded to the oxygen atom. This discrepancy of 0.22 eV in LP could be due to the different cell sizes used in our calculations. At site-2, the Li atom remains at the hollow site and hence has a lower LP as expected. We also looked at adsorption of Li atom at site-3,4,5 on the same side as epoxy and found that in all three cases, Li atom gets bonded with the oxygen atom resulting in 1.44 eV of LP(same as site-1 conf. Figure[\ref{fig32}](II)), with O-Li bond getting oriented along the hollow site. Thus, LP gets greatly affected whether the Li atom is deposited on the same side or the opposite side of the epoxy group in the sheet. For the case of two Li atoms in cell at site-1\&2 and site-1\&4 placed along the side of epoxy, we obtain a LP of 1.11 eV and 2.01 eV, respectively. For site-1\&2 one Li atom makes bond to oxygen and another Li atom remaining at site-2, while for site-1\&4 Li atoms come near to oxygen with the position of oxygen at onsite position shown in Figure[\ref{fig32}](III). 

In the case hydroxyl functionalized graphene sheet, we get the highest LP of 0.86 eV at site-6 and 7 with Li atom deposited on the opposite side of the -OH group. Thus, LP in the hydroxyl functionalized graphene is higher than that of the epoxy functionalized graphene. With two and three Li atoms adsorbed on the opposite side, LP of 0.32 eV and 0.22 eV (specific capacity  230 \mahg and  345 \mahg), respectively are obtained. In the extreme case of fully covered (nine Li atoms per -OH group) GO a LP of -0.20 eV is found. However when Li is placed on the same side as the  -OH group, it gets strongly bonded to the -OH group and the whole molecule (Li-OH) gets detached from graphene sheet.

\begin{figure}[h]
\centering
\includegraphics[width=78mm, height=59mm]{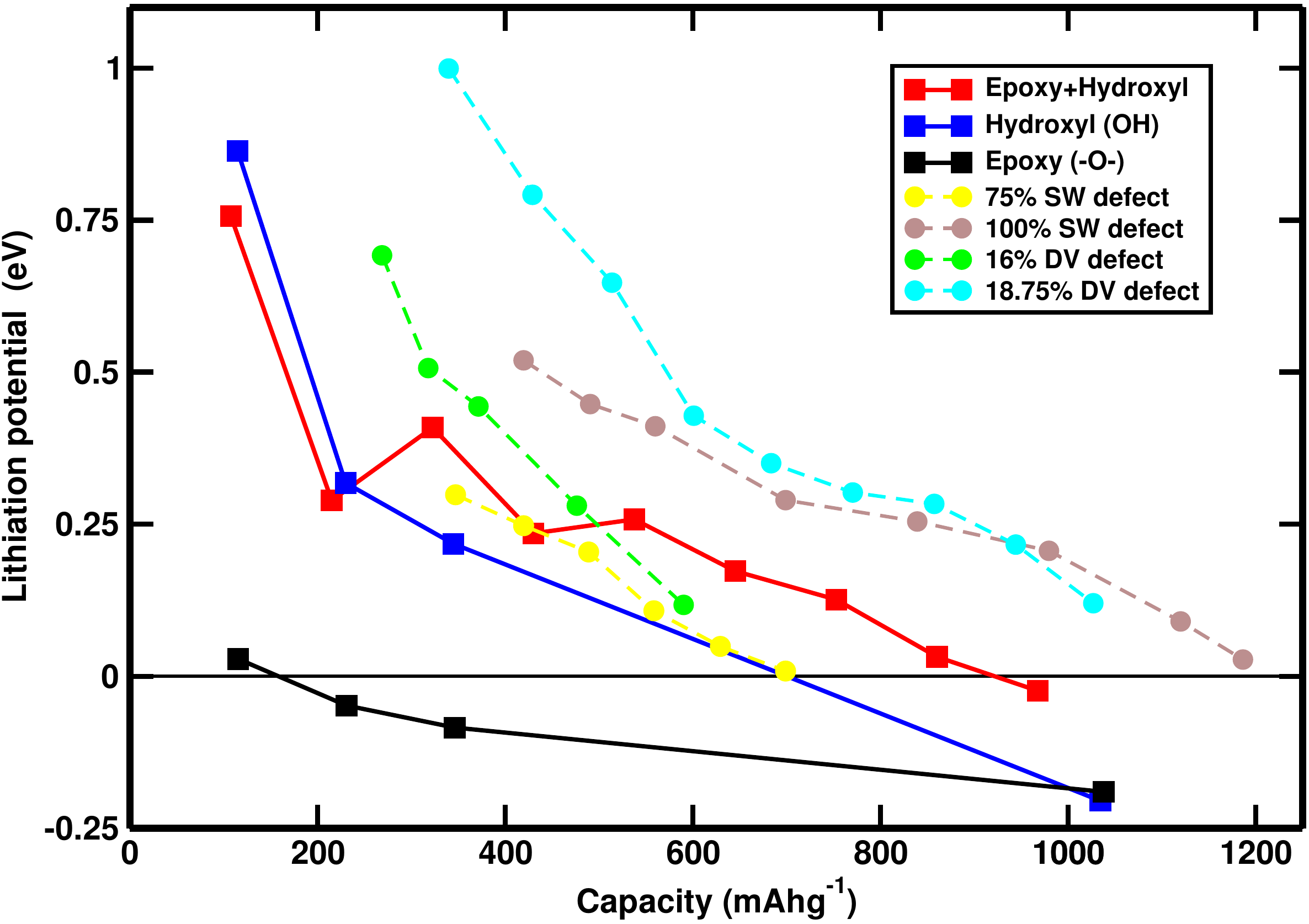}
\caption {Lithiation potential as a function of capacity for epoxy, hydroxyl, and both (-O \& -OH) functional groups attached to the graphene sheet and compared with others.\cite{mukherjee2014,dutta2014} Squares represent LP with functional groups, while circles represent data from previous literation with SW and DV defects in graphene sheet. A highest capacity of $\sim$860 \mahg is obtained on GO with both epoxy and hydroxyl groups attached.}
\label{fig4}
\end{figure}

We then investigated the LP when both groups (epoxy and -OH) are presents in the graphene sheet and placed on the same side. The possible orientations of -OH with respect to epoxy are shown in Figure[\ref{fig3}](III). Out of (a)-(h) configurations, the lowest energy configurations are found to be (e) and (h). Lithium atoms are place on the opposite side of the functional groups. LP for one Li in 3$\times$3 supercell (107 \mahg) is 0.76 eV, which is 0.1 eV less than that obtained with only -OH group. However, as we gradually increase the concentration of Li, LP increases compared to that with only OH group, as shown in Figure[\ref{fig4}] (compare red squares with blue squares), \textit{e.g.}, LP at 315 \mahg is 0.19 eV  greater than that with only OH group. The close analysis of the structure shows that, above 315 \mahg capacities, epoxy group breaks one of its bond and oxygen gets attached to a single C atom at onsite position in the sheet. We observe a similar behaviour when epoxy contained graphene sheet was fully covered with Li on the opposite side (average LP = -0.19 eV) as shown in Figure[\ref{fig32}](IV). This new behaviour of oxygen atom replicates a situation similar to that of the hydroxyl group that shows a higher LP compared to the epoxy group. This breaking of the bond of oxygen with one of the carbon atoms could be caused by the stretching of the graphene sheet in presence of the metallic Li layer on the opposite side. Charge redistribution in presence of Li atoms (particularly the 2s orbital contribution) and the singly bonded oxygen atom with graphene sheet at onsite position could be important factors for the boost in the LP with higher capacity. From Table[\ref{table2}], notice that the configuration III(e) dominates the binding upto the capacity of 645 \mahg with LP of 0.17 eV, while from 752 \mahg capacity onwards configuration III(h) gives the highest capacity $\sim$860 \mahg with LP of 0.03 eV. Thus, we successfully achieve a positive LP upto the specific capacity of $\sim$860 \mahg just by placing the Lithium atoms only on one side of the sheet as shown in Figure[\ref{fig32}](V) \& (VI).

\begin{table}
\small
\centering
\caption{Lithiation potential with varying capacity for conf. III(e) and III(h) in 3$\times$3 supercell.}
\label{table2}
  \begin{tabular}{ccccc}\hline \hline
    Capacity   &  Conf. III(e)  & Conf. III(h) & \# of Li \\
  (\mahg)       & eV          & eV       & atoms  \\ \hline
 107         &   0.76      & 0.54   &  1 \\        
 215         &   0.29      & 0.26   &  2 \\   
 322         &   0.41      & 0.28   &  3 \\  
 430         &   0.24      & 0.20   &  4 \\
 537         &   0.26      & 0.12   &  5 \\
 645         &   0.17      & 0.16   &  6 \\
 752         &   0.08      & 0.13   &  7 \\
 860         &   0.02      & 0.03   &  8 \\      
 967         &  -0.09      & -0.02  &  9 \\   \hline
\end{tabular}
\end{table}

\section{Conclusions}

Lithium adsorption and lithiation potential on graphene oxide has been investigated using DFT. We find that epoxy and hydroxyl functionalized graphene sheet provides the highest specific capacity of $\sim$860 \mahg. We have achieved higher capacities than previously reported capacities\cite{dutta2014} of 698 \mahg and 590 \mahg with 75\% SW and 16\% DV defects. Although, even higher capacities of 1186 \mahg and 1029 \mahg were obtained by Dutta \etal,\cite{dutta2014} with 100\% SW and 18.75\% (or higher) DV defects, respectively. We believe that creating such high density of defects would make the entire system energetically unstable leading to bending and wrapping of the 2D sheet. In this paper, we have shown the hike in Li storage capacity by attaching the functional group on one side of the sheet and Li on the other side. However, it would be possible to further increase the storage capacity by increasing the density of functional groups and Li atoms, \textit{i.e.}, by playing around with number of Li atoms per epoxy and/or per hydroxyl groups. We have also demonstrated that London's dispersion contribution plays an important role in binding of Li on graphene and GO sheets. Oxygen bonding and its position in GO plays an important role in raising the specific capacity of GO. Thus, based on the first principle studies, we propose an efficient way to increase the capacity on a novel low-cost system (GO). Our results provide an insight to find a better anode material for rechargeable Li batteries without damaging or drastically modifying the properties of graphene.

\section{Acknowledgement}
The authors thank the HPC center of Idaho National Laboratory for providing computational facilities. This work was financially supported from NSF CAREER award (DMR-1255584) and Research Corporation's Cottrell college Science award.
%%%%%%%%%%%%%%%%%%%%%%%%%%%%%%%%%%%%%%%%%%%%%%%%%%%%%%%%%%%%%%%%%%%%%
%                     Reference Page
%%%%%%%%%%%%%%%%%%%%%%%%%%%%%%%%%%%%%%%%%%%%%%%%%%%%%%%%%%%%%%%%%%%%%%


\begin{thebibliography}{99}
\bibitem{howel2008}D. Howell, T. Duong, J. B. Deppe, I. Weinstock,  \textit{Material Matters }, 2008, \textbf{3.4}, 100.
\bibitem{ref2}G-. A. Nazri, G. Pistoia, \textit{Springer }, 2009, \textbf{ISBN: 978-0-387-92675-9}.
\bibitem{muzushima1980}K. Mizushima, P. C. Jones, P. J.  Wiseman, J. B. Goodenough,  \textit{Materials Research Bull}, 1980, \textbf{15}, 783-789.
\bibitem{tarascon2001}J-. M. Tarascon, M. Armand, \textit{Nature}, 2001, \textbf{414}, 359-367.
\bibitem{bhattacharya2010} R. Bhattacharyya,	B. Key, H. Chen, A. S. Best, A. F. Hollenkamp, C. P. Grey, \textit{Nature Materials}, 2010, \textbf{9}, 504-510.
\bibitem{orsini1998}F. Orsini, A. D. Pasquier, B. Beaudoin, J. M. Tarascon, M. Trentin,N. Langenhuizen, E. D. Beer, P. Notten, \textit{J. Power Sources},  1998, \textbf{76}, 19-29.
\bibitem{mukherjee2014}R. Mukherjee, A. V. Thomas, D. Datta, E. Singh, J. Li, O. Eksik,V. B. Shenoy,N. Koratkar, \textit{Nature Communications},  2014, \textbf{5}, 3710.
\bibitem{Liu2012}N. Liu, H. Wu, M. T. McDowell, Y. Yao, C. Wang, Y. Cui, \textit{Nano Letters},  2012, \textbf{12}, 3315-3321.
\bibitem{Zhou2012}X. Zhou,Y- X. Yin,L- J.  Wan, Y- G. Guo,  \textit{Chemical Commun.}, 2012, \textbf{48}, 2198-2200.
\bibitem{Krishnan2011} R. Krishnan,T- M. Lu, N. Koratkar, \textit{Nano Letters}, 2011, \textbf{11}, 377-384.
\bibitem{Stournara2011}M. E.  Stournara,V. B.  Shenoy, \textit{J. of Power Sources}, 2011, \textbf{196}, 5697-5703.
\bibitem{Meunier2002}V. Meunier, J. Kephart,C. Roland, J. Bernholc, \textit{Phys. Rev. Lett.}, 2002, \textbf{88}, 075506.
\bibitem{dutta2014}D. Dutta, J. Li, N. Koratker, V. B. Shenoy, \textit{Carbon}, 2014, \textbf{80}, 305-310.
\bibitem{li2005}L. Li, S. Reich, J. Robertson, \textit{Phys. Rev. B}, 2005, \textbf{72}, 184109.
\bibitem{ma2009}J. Ma, D. Alfe, A. Michaelides, E. Wang,  \textit{Phys. Rev. B}, 2009, \textbf{80}, 033407.
\bibitem{banhart2011}F. Banhart, J. Kotakoski, A. V. Krasheninnikov, \textit{ACS Nano}, 2011, \textbf{5}, 26-41.
\bibitem{krasheninnikov2006}A. V. Krasheninnikov, P. O. Lehtinen, A. S. Foster, R. M. Nieminen, \textit{Chem. Phys. Lett.}, 2006, \textbf{418}, 132-136.
\bibitem{rossato2005}J. Rossato, R. J. Baierle, A. Fazzio, R. Mota, \textit{Nano Lett.}, 2005, \textbf{5}, 197-200.
\bibitem{barbary2003}A. A. El-Barbary, R. H. Telling, C. P. Ewels, M. I. Heggle, P. R. Briddon, \textit{Phys. Rev. B},  2003, \textbf{68}, 144107.
\bibitem{loh2010}K. P. Loh, Q. Bao, G. Eda, M. Chhowalla, \textit{Nature Chem.}, 2010, \textbf{2}, 1015.
\bibitem{suk2010}J. W. Suk, R. D. Piner, J. An, R. S. Ruoff,  \textit{ACS Nano}, 2010, \textbf{4}, 6557-6564.
\bibitem{hummers1958} R. E. Offeman, W. S. Hummers, \textit{J Am Chem Soc}, 1958, \textbf{80}, 1339-1339.
\bibitem{marcano2010}D. C. Marcano \etal,   \textit{ACS Nano}, 2010, \textbf{4}, 4806-4814.
\bibitem{dreyer2010}D. R. Dreyer, S. Park, C. W. Bielawski, R. S. Ruoff, \textit{Chem. Soc. Rev.}, 2010, \textbf{39}, 228-240.

\bibitem{QE}P. Giannozzi \etal, \textit{J. Phys.: Condens. Matt.}, 2009, \textbf{21}, 395502.
\bibitem{perdew1996}J. P. Perdew, K. Burke, M. Ernzerhof, \textit{Phys. Rev. Lett.}, 1996, \textbf{77}, 3865.
\bibitem{Allouche2012} A. Allouche, P. S. Krstic, \textit{Carbon}, 2012, \textbf{50}, 510-517.
\bibitem{grimme2006}S. Grimme, \textit{J. Comp. Chem.}, 2006, \textbf{27}, 1787-1799.
\bibitem{tkatchenko2009}A. Tkatchenko, M. Scheffler, \textit{Phys. Rev. Lett.}, 2009, \textbf{102}, 073005.
\bibitem{marzari1999}N. Marzari, D. Vanderbilt, A. de Vita, M. C. Payne, \textit{Phys. Rev. Lett.}, 1999, \textbf{82}, 3296.
\bibitem{trucano1975}P. Trucano, R. Chen, \textit{Nature}, 1975, \textbf{258}, 136-137.
\bibitem{Aydinol1996}M. K. Aydinol, A. F. Kohan, G. Ceder, \textit{J Power Sources}, 1997, \textbf{68}, 664-668.

\bibitem{chan2008} K. T. Chan, J. B. Neaton, M. L. Cohen, \textit{Phys. Rev. B}, 2008, \textbf{77}, 235430.
\bibitem{Garay-Tapia2012} A. M. Garay-Tapia, A. H. Romero, V. Barone, \textit{J. Chem. Theory Comput.}, 2012, \textbf{8}, 1064-1071.
\bibitem{Zhou2012_2}L- J. Zhou, Z. F. Hou, L- M  Wu, \textit{J. Phys. Chem. C}, 2012, \textbf{116}, 21780-21787.
\bibitem{Robledo2014}C. B. Robledo, M. Otero, G. Luque, O. Camara, D. Barraco, M. I. Rojas, E. P. M. Leiva, \textit{Electrochimica Acta}, 2014, \textbf{140}, 232-237.
\bibitem{cai2008} W. Cai \etal, \textit{Science}, 2008, \textbf{321}, 1815-1817.
\bibitem{zhou2013}S. Zhou, A. Bongiorno, \textit{Nature Scientific Rep.}, 2013, \textbf{3}, 2484.
\bibitem{wang2009}G. Wang, J. Yang, J. Park, X. Gou, B. Wang, H. Liu, J. Yao, J. \textit{J. Phys. Chem. C}, 2008, \textbf{112}, 8192-8195.
\bibitem{staudenmaier1898}L. Staudenmaier, \textit{Ber. Dtsch. Chem. Ges.}, 1898, \textbf{31}, 1481-1487.
\bibitem{stanko2007}S. Stankovich \etal,  \textit{Carbon}, 2007, \textbf{45}, 1558-1565.

\end{thebibliography}
\end{document}